\documentclass[sigconf]{acmart}

\usepackage{float}
\usepackage{graphicx}
\AtBeginDocument{%
  \providecommand\BibTeX{{%
    \normalfont B\kern-0.5em{\scshape i\kern-0.25em b}\kern-0.8em\TeX}}}

\copyrightyear{2021}
\acmYear{2021}
\setcopyright{acmlicensed}\acmConference[CHI '21 Extended Abstracts]{CHI Conference on Human Factors in Computing Systems Extended Abstracts}{May 8--13, 2021}{Yokohama, Japan}
\acmBooktitle{CHI Conference on Human Factors in Computing Systems Extended Abstracts (CHI '21 Extended Abstracts), May 8--13, 2021, Yokohama, Japan}
\acmPrice{15.00}
\acmDOI{10.1145/3411763.3451663}
\acmISBN{978-1-4503-8095-9/21/05}

\begin{document}

\title{Older Adults and Brain-Computer Interface: An Exploratory Study}


\author{Wiesław Kopeć}
\email{kopec@pjwstk.edu.pl}
\orcid{0000-0001-9132-4171}
\affiliation{%
  \institution{Polish-Japanese Academy of Information Technology}
  \city{Warsaw}
  \country{Poland}
}

\author{Jarosław Kowalski}
\email{jaroslaw.kowalski@opi.org.pl}
\orcid{0000-0002-1127-2832}
\affiliation{%
  \institution{National Information Processing Institute}
  \city{Warsaw}
  \country{Poland}
}

\author{Julia Paluch}
\email{jpaluch@pjwstk.edu.pl}
\orcid{0000-0002-7657-7856}
\affiliation{%
  \institution{Polish-Japanese Academy of Information Technology}
  \city{Warsaw}
  \country{Poland}
}

\author{Anna Jaskulska}
\email{a.jaskulska@kobo.org.pl}
\orcid{0000-0002-2539-3934}
\affiliation{%
  \institution{Kobo Association}
  \city{Warsaw}
  \country{Poland}
  }

\author{Kinga Skorupska}
\email{kinga.skorupska@pjwstk.edu.pl}
\orcid{0000-0002-9005-0348}
\affiliation{%
  \institution{Polish-Japanese Academy of Information Technology}
  \city{Warsaw}
  \country{Poland}
  }

\author{Marcin Niewiński}
\email{marcin.niewinski@pjwstk.edu.pl}
\orcid{0000-0002-6416-3541}
\affiliation{%
  \institution{Polish-Japanese Academy of Information Technology}
  \city{Warsaw}
  \country{Poland}
  }
  
\author{Maciej Krzywicki}
\email{mkrzywicki@pjwstk.edu.pl}
\orcid{ }
\affiliation{%
  \institution{Polish-Japanese Academy of Information Technology}
  \city{Warsaw}
  \country{Poland}
  }
  
  \author{Cezary Biele}
\email{cezary.biele@opi.org.pl}
\orcid{0000-0003-4658-5510}
\affiliation{%
  \institution{National Information Processing Institute}
  \city{Warsaw}
  \country{Poland}
}


\renewcommand{\shortauthors}{Kopeć et al.}

\begin{abstract}In this exploratory study, we examine the possibilities of non-invasive Brain-Computer Interface (BCI) in the context of Smart Home Technology (SHT) targeted at older adults. During two workshops, one stationary, and one online via Zoom, we researched the insights of the end users concerning the potential of the BCI in the SHT setting. We explored its advantages and drawbacks, and the features older adults see as vital as well as the ones that they would benefit from. Apart from evaluating the participants' perception of such devices during the two workshops we also analyzed some key considerations resulting from the insights gathered during the workshops, such as potential barriers, ways to mitigate them, strengths and opportunities connected to BCI. These may be useful for designing BCI interaction paradigms and pinpointing areas of interest to pursue in further studies.

\end{abstract}

\ccsdesc[500]{Human-centered computing~Interaction paradigms}
\ccsdesc[500]{Social and professional topics~Seniors}
\ccsdesc[100]{Human-centered computing~HCI design and evaluation methods}
\ccsdesc[300]{Human-centered computing~Ubiquitous and mobile devices}

\keywords{BCI, brain-computer interface, IoT, internet of things, older adults, smart home, living lab}

\maketitle

\section{Introduction}

Due to demographic ageing\cite{united_nations_department_of_economic_and_social_affairs_world_2020}
and its socioeconomic consequences \cite{socioeconageingcon2014}, addressing the needs of older adults is becoming a priority. 
One of the rapidly developing technologies which may be utilised in this context is non-invasive brain-computer interaction (BCI), which introduces new possibilities of creating devices and systems aimed at older adults.\cite{belkacem_brain_2020} Together with Smart Home Technology, BCI can improve their quality of life by accommodating their needs and removing barriers caused by some age-related health issues\cite{jaul_age-related_2017, chang_measuring_2019} such as restrained mobility, which can happen along with other causes of non-fatal health loss as people continue to live longer\cite{USAolderhealth2019}. For such users BCI can empower them to benefit from ICT solutions which otherwise might be partially inaccessible to representatives of their age group. Existing studies confirm the feasibility of utilizing BCI as an interface approachable and beneficial for older adults. 

In our study, we set out with a goal of examining the topic of brain-computer interaction from a different perspective: to gain insights directly from the target group. Our objective was to explore the viewpoint of older adults themselves - how they perceive the BCI technology, whether they would be willing to use it in a Smart Home setting, and what would be their expectations towards the capabilities of such a system. We wanted to put emphasis on the needs of potential end users, whose insight could be beneficial for designing BCI-based Smart Home interaction paradigms.

\section{Related Work}
Previous studies confirmed the feasibility of using BCI technologies in a medical setting to aid elderly patients with multiple health issues \cite{belkacem_brain_2020}, such as post-stroke rehabilitation \cite{foong_assessment_2020}, disorders of consciousness \cite{xiao_visual_2018}, and motor system issues (allowing them to operate a wheelchair \cite{kaufmann_toward_2014} and a robotic knee exoskeleton \cite{herweg_wheelchair_2016}). Recently, researchers have shown a rising interest in utilizing BCI in Smart Home technology targeted at older users (as in Jafri et al.\cite{jafri_wireless_2019} and Chai et al.\cite{chai_hybrid_2020}). This direction promises a spectrum of possibilities for making a positive impact on older adults' lifestyle, accessibility to technology, and managing the symptoms of the ailments of old age. There are multiple exploratory studies on older adults' perception of new technologies which produced valuable insights for the industry, either based on workshops \cite{kowalski_older_2019}, surveys \cite{acceptanceoftechnologysurv2005} or data models. \cite{datamodeloldertech2007} 

The studies on the older adults' perception of smart home systems date to over a decade back, as seen on the example of Demris et al.\cite{demiris_older_2004} and Sarkisian et al.\cite{sarkisian_older_nodate}. There are also studies on the topic of BCI interaction conducted with other age groups, for example, with high school students in a study by Hernandez-Cuevas et al. \cite{bcihighschool2020chi}. However, there is a clear need to further explore the niche of older adults' attitude towards using brain-computer interaction in the context of controlling smart home and IoT devices. 

\section{Methods}
We invited seven older adults who were previously involved in our LivingLab activities to participate in the study. The group consisted of four female and three male participants, aged from 60 to 80, and came from a city with a population of over 1 million. 
They were retired from different occupational backgrounds. None of them had serious mobility or cognitive issues and all of them gave their informed consent to take part in the study.
In the selection of the group we were guided by making it as diverse as possible - it consisted of people of both genders, of different ages, and with different levels of familiarity with electronic devices such as computer, tablet, or smartphone. However, it is worth noting that their common trait was their interest in new technologies, and all had above average ICT skills for this age group. They were active socially and intellectually and engaged in multiple activities such as presence in the local citizen community forums or previous participation in ICT-oriented workshops, for example on VR or Smart Home devices; thereby granting them the ability to refer to these technologies in connection with BCI.

The study was conducted in two separate sessions, each lasting approximately for two hours. The first session was held onsite, while the second took place online on the Zoom video conference platform. Semi-structured group interviews were utilized for both sessions, the outline of which is described below. The content of the interviews was later transcribed and annotated for further inductive and deductive thematic analysis.\\

\begin{figure}[H]
  \centering
    \includegraphics[width=0.7\linewidth]{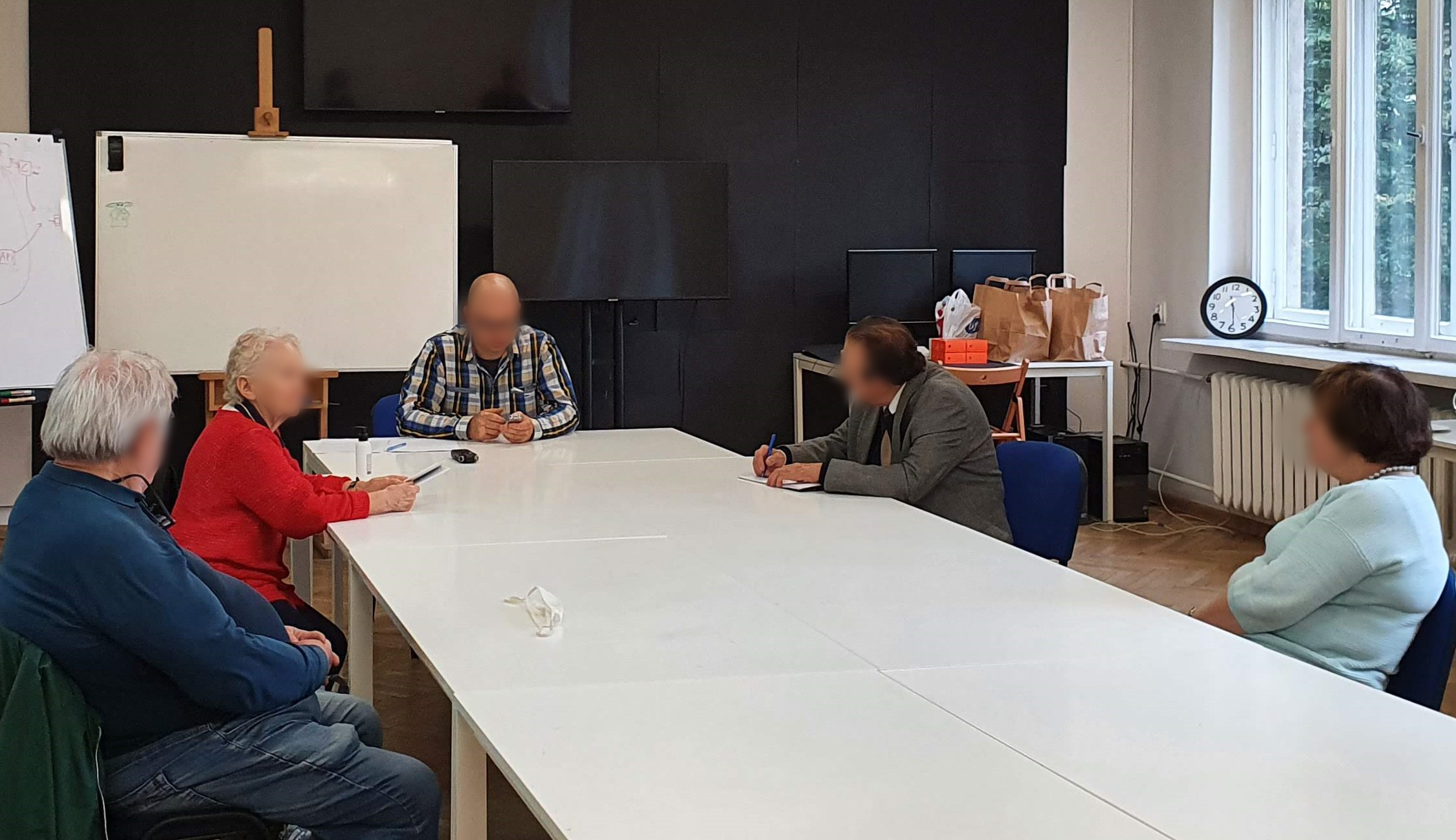}
  \caption{The workshop was based on a semi-structured qualitative scenario. The participants were encouraged to share their insights and associations at all times, and the subsequent questions were modified dynamically to adhere to the natural flow of the conversation.}
  \Description{The workshop was based on a semi-structured qualitative scenario. The participants were encouraged to share their insights and associations at all times, and the subsequent questions were modified dynamically to adhere to the natural flow of the conversation.}
  \label{workshoppiccie}
\end{figure}

The semi-structured study scenario consisted of 6 parts:
\begin{enumerate}
\item \textbf{Free association:} exploring the first associations and observations of the participants about BCI in a Smart Home setting,
\item \textbf{Brainstorming:} discussing the potential use of brain-computer interfaces in everyday activities,
\item \textbf{Showcasing:} introducing the concept of headband acting as a BCI controller (based on monitoring the electric activity in the brain) and collecting the participants' insights on the technology,
\item \textbf{Comparing:} the participants are asked to compare and contrast brain-computer interfaces and voice interfaces as based on Google Home voice assistant,
\item \textbf{Verifying:} collecting the opinions on whether the participants would be willing to try headband-based BCI solutions in their own home,
\item \textbf{Analysing:} discussing potential opportunities and threats associated with using BCI in Smart Home technology.
\end{enumerate}

\section{Results}
In general, we observed that most of the participants were open-minded and intrigued by the idea of brain-computer interface. They perceived it as modern and advanced. However, they mostly remained cautious and hesitant to use the technology in a smart home setting without being well-informed about its principles. Many did not see an advantage in using it instead of a Smart Home system. At this point of the development of the technology, they perceived voice assistants as superior and lacking some of the drawbacks of BCI.
The most emphasized aspect of the potential use of BCI was in healthcare and emergency response, which was unanimously agreed on during both workshop sessions.\\
Below we present the most noteworthy insights which emerged during each part of the interview scenario:

\begin{enumerate}
\item \textbf{Free association:} \\
• Overall, the participants displayed a positive attitude towards the idea of using BCI in smart home systems. However, it should be noted that they were a group of older adults with above basic ICT skills, most of them familiar with using computers or smartphones.\\
• Notably, the older adults taking part in the study showed remarkable understanding and support for developing new technologies such as brain-computer interfaces.
• As P3 stated, "using such technology is an attractive challenge, making everyday life easier and leaving more free time and space to focus on personal goals".\\
• One participant (P5) said the technology seems "futuristic and unreal". Other discussed how elusive thoughts are and whether it is possible to use them in IT purposes.\\

\item \textbf{Brainstorming:}\\
• In one of the workshop groups participant (P2) suggested the potential use of BCI in the context of emergency situations; the ensuing discussion produced valuable insights. All participants expressed their enthusiasm for the idea of using the headband for responding to emergency
situations and automatically alerting the relevant medical services (such as paramedics). Participants (P1, P2, P4, P5) were able to cite examples of situations from their lives (or those of their friends) where such a solution could have saved someone's health or life. They listed a number of advantages, such as the ability to communicate with the device in the event of immobility, loss of consciousness, shock, or disorientation. They pointed out how such functionality would be useful especially for the elderly, suffering from chronic illnesses, or living alone (who have no one to help them in case of an accident). \\
• One of the participants (P2) proposed to use the capability of recognizing thought patterns to diagnose illnesses affecting thinking, such as psychiatric disorders. This notion was also positively received by the rest of the group.\\
• In addition to its medical potential, it could also support crime prevention because, in addition to automatic alerting, it would make it possible to notify the authorities in circumstances requiring discretion in order not to alert an assailant. As P3 pointed out, the safety measures are of the utmost importance and should be prioritized; solutions aimed at improving the comfort of the user can be developed next.\\

\item \textbf{Showcasing:}\\
 • When informed that the headband will recognize fixed thought patterns, the participants voice several doubts about the solution. They point out that it will require to learn fixed, "stiff" thought patterns and constantly remember to use them. For some, it means the interaction will be less natural - more mechanical and less spontaneous, "the way that robots talk". It is worth noting since multiple participants said they would wish the communication with the interface was effortless and natural, like talking to an other human. Some participants (P1, P2) mentioned that they wish the BCI was programmed to have a sense of humor and displayed its own personality traits (warm, friendly and occasionally witty), similar to voice assistants such as Alexa or Siri (which can be asked to tell a joke or sing, for instance).\\
• The notion of the BCI recognizing predefined, fixed thought patterns was discouraging for some, similar to the necessity to use the headband to access the smart home system. For other users it might be tiring to constantly remember to use fixed thought patterns that are recognized by the interface. As one participant put it, it could lead to a situation when "you must think in a certain way or you won't be able to live in a smart home".\\

\item \textbf{Comparing:} \\
• The participants pointed out several factors contributing to the perceived inferiority of BCI to voice assistants. One of the most criticized elements was the headband, as the participants voiced doubts about the necessity to constantly carry it along, remember to put it on their head, and the general lack of comfort. It was noted that a voice assistant could perform the same tasks without added drawbacks. Furthermore, it distinguishes different voices easily and speaks in a manner that seems more approachable and natural, similar to talking to other person.
On the other hand, it was pointed out that the brain computer interfaces would allow to discreetly give commands without alerting or disturbing people nearby. It would also be accessible to users with speech impairments.\\
• In older interfaces, an essential stage of interaction was external action (pushing a button, clicking an icon, voicing a command in case of voice interfaces), causing an effect. Brain-computer interfaces omit this stage;
a command is issued using only the brain activity:\\
\begin{figure}[H]
  \centering
  \includegraphics[width=\linewidth]{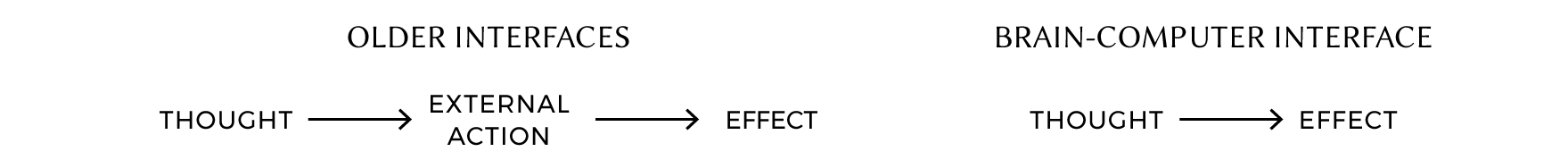}
  \caption{Approximation of the outline of interaction in different interfaces.}
  \Description{Approximation of the outline of interaction in different interfaces.}
  \label{cards}
\end{figure}
One of the potential threats that emerged during the workshops is the fact that omitting the decision to voice a command removes a natural verification opportunity; the system would need another verification solution to ensure no commands are given by accident, for instance when considering a task or thinking about it in a different context. \\

\item \textbf{Verifying:}\\
• The participants (P1, P5, P6) emphasized the impact that smart home BCI could have on improving the life quality of people with chronic illnesses, with mobility issues, speech or hearing impairments, or depressed and unable to perform tasks on their own. This purpose was significantly prioritized, while other applications were mostly considered a fun novelty (or means to save time and effort) which could potentially be developed into new standardised way of communicating with electronic devices.\\
• Multiple participants (P2, P7) voiced their concerns about the price of the device. They were apprehensive that, due to the novelty and level of advancement of the BCI technology, it would not be affordable for an average user and, therefore, would not come into widespread use.\\
• Notably, some of the participants were eager to test the technology for the sole purpose of contributing to scientific development. (P7) stated that she "would be happy if the data collected from her could help scientists understand how the human brain works," while (P6) said that every new invention is an opportunity to extend our knowledge and to accelerate scientific research and, therefore, should be supported.\\
• The participants noticed the potential of using BCI in a Smart Home environment to save time and effort while performing everyday activities and some were willing to use it if it made routine tasks "easier or more attractive" (P7).\\

\begin{figure}[H]
  \centering
  \includegraphics[width=\linewidth]{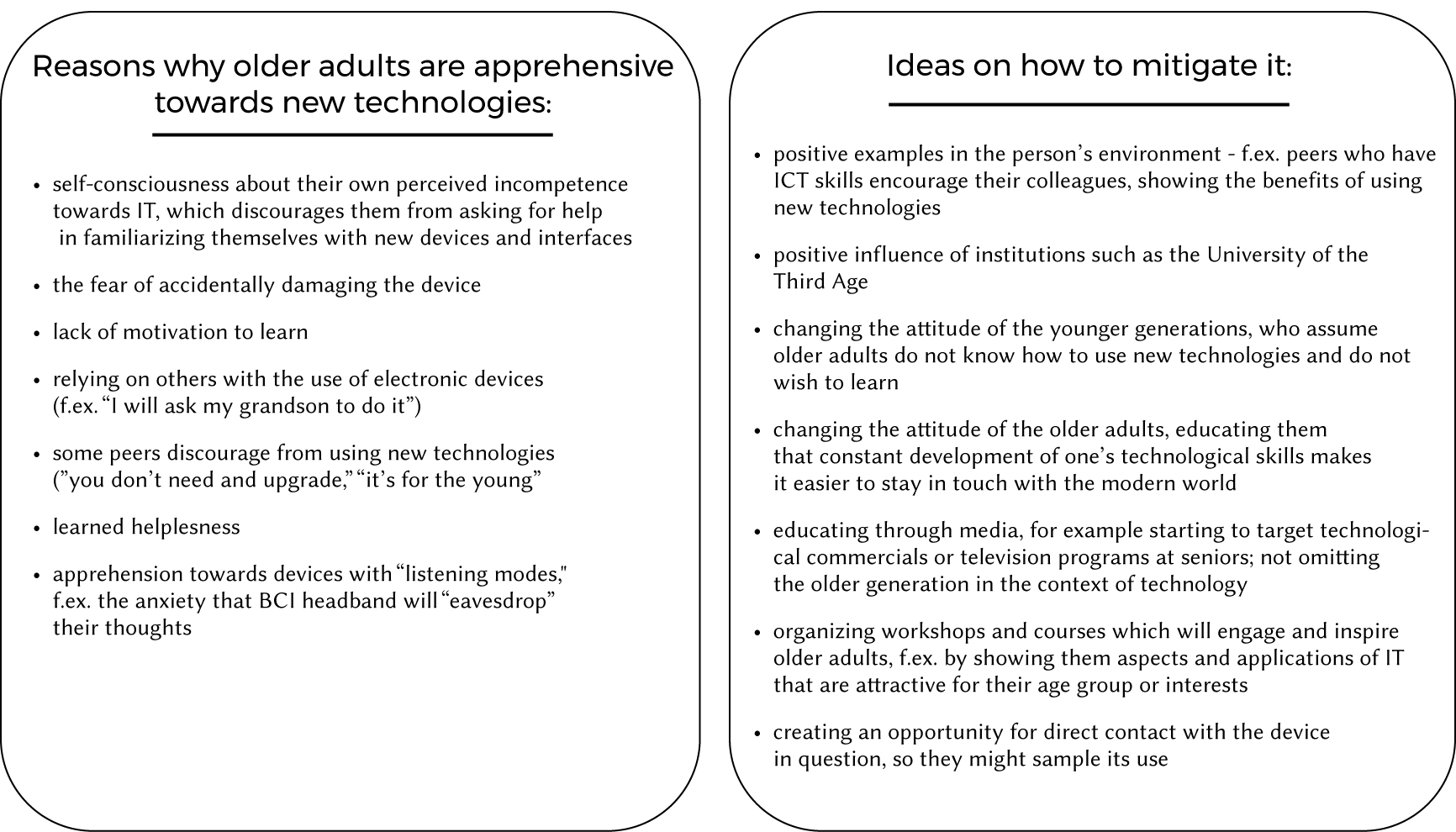}
  \caption{Insights from the workshop participants' comments about older adults' approach to new technologies, such as BCI.}
  \Description{Insights from the workshop participants' comments about older adults' approach to new technologies, such as BCI.}
  \label{reasonsideas}
\end{figure}

\item \textbf{Analysing:}\\
• The participants (most notably P3) voiced concerns about the use of the device by people with dementia, psychiatric issues, or thinking emotionally; when they think in a chaotic manner, their thought might be misinterpreted or cause a task which might be dangerous to them and their surroundings (e.g. "turn on the stove"). One participant (P3) suggested there should be a feature verifying whether the user is capable of using the system responsibly, for instance if they are not under the influence of psychoactive substances.  \\
• Lack of privacy and the danger of a third party gaining access to the data stored by the smart home technology (which is especially stressful considering personal and intimate character of a person's thoughts) was an issue to some. However, as stated above, some would not mind the data collecting if its analysis would be beneficial for science.\\
• As described above, all participants reacted enthusiastically to the notion of using BCI in medical setting and for emergency response. They also emphasized its potential in Smart Home Technology for users who have mobility issues, speech impairments, are depressed, or simply live at a fast pace and want to save time.\\

\end{enumerate}

\textbf{The key needs reported by the potential users:}
\begin{enumerate}
\item \textbf{} the capability to access and send commands to electronic devices and appliances used in everyday activities (TV, PC, fridge, lightning)
\item \textbf{} a "central switch" allowing to turn off all electronic devices at home at once, for example to ensure that potentially dangerous appliances such as a clothes iron are unplugged
\item \textbf{} notifications to confirm whether a command was understood by the device or to inform when a task has been finished
\item \textbf{} a possibility to gain remote access to the smart home system
\item \textbf{} safety features activating in emergency situations, e.g. during an accident or a stroke
\end{enumerate}

\section{Discussion}

\begin{figure}[H]
  \centering
  \includegraphics[width=\linewidth]{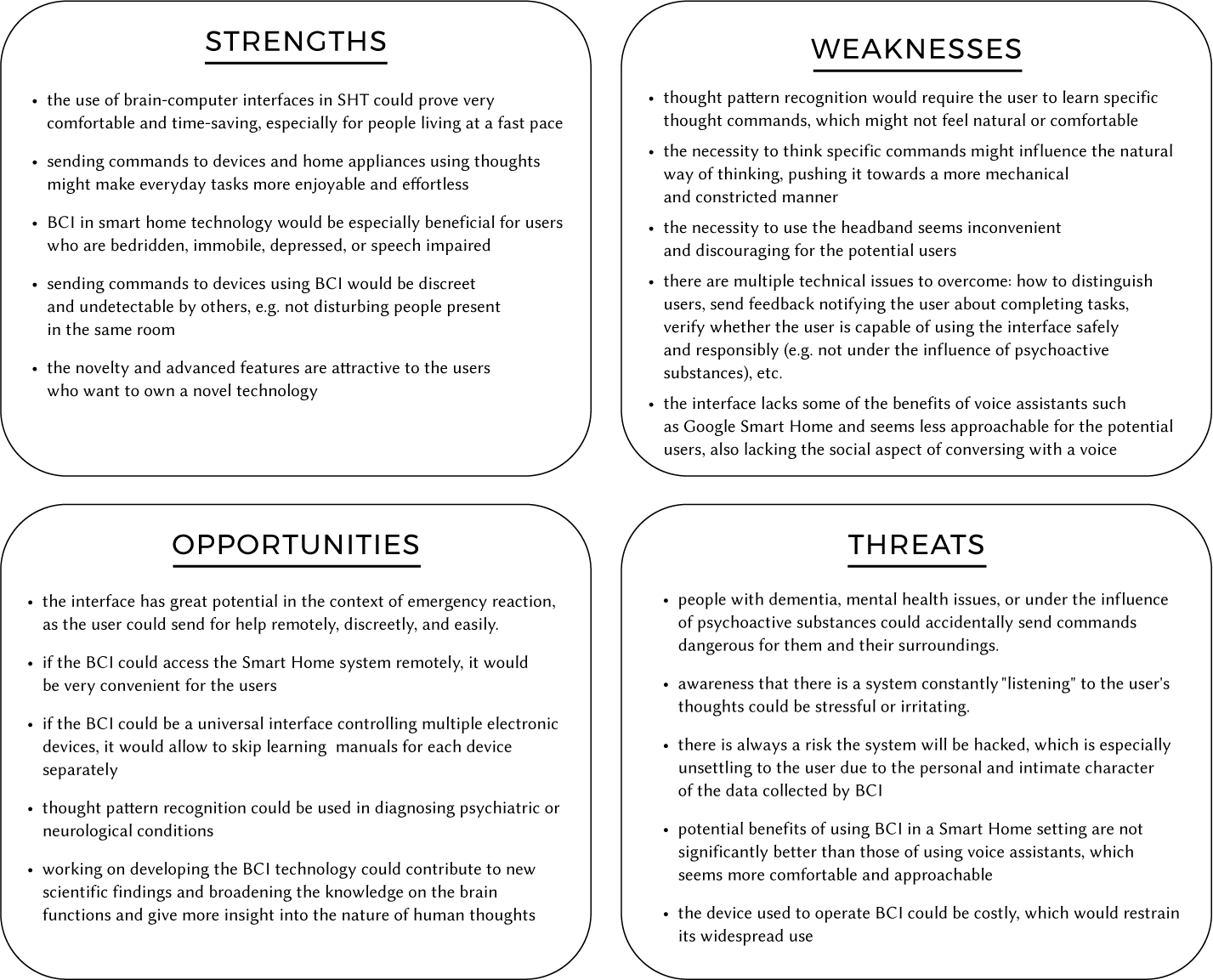}
  \caption{SWOT analysis of the brain-computer interfaces in the context of the Smart Home Technology based on our findings from the workshops with older adults.}
  \Description{SWOT analysis of the brain-computer interfaces in the context of the Smart Home Technology based on our findings from the workshops with older adults.}
  \label{swot}
\end{figure}

Conducting the workshop allowed us to achieve the following objectives:
\begin{enumerate}
\item explore the older adults' attitudes towards and perception of brain-computer interfaces in the context of Smart Home technology, 
\item identify their key needs as users,
\item research the advantages and disadvantages of the technology as seen by the older adults,
\item familiarize with their outlook on the introduced concept of headband-based BCI device,
\item verify whether they would be willing to use the technology in their own homes.
\end{enumerate}
Moreover, during the interviews, we gained valuable insights on the ideas for the potential use of BCI and confirmed our assumption that their age group's attitude towards new technologies can be positive in general, given the right context in which the technologies are introduced. Some ideas on how to achieve the same positive results by addressing barriers to ICT-use for more people from this age group are shown in Fig. \ref{reasonsideas}. Moreover, in Fig. \ref{swot} we present a SWOT analysis for the introduction of BCI technology in the context of older adults, which may guide implementation attempts for this age group and may serve as a guideline in planning further research in this area.
We observed that a small number of participants (3-4) in each workshop session was appropriate for our interview scenario, as it allowed each person to voice their opinions and actively engage in the discussion. However, the small size of the group should be considered a limitation of our study; it would be beneficial to repeat our study with more participants. An added advantage would also be including those with basic or no ICT skills, as well as older adults with mobility issues and other health problems associated with advanced age.

\section{Conclusions}
\balance
The conclusions we have drawn from the study, which are summarized in Fig. \ref{swot}, confirmed our assumption that some groups of older adults, especially those more proficient in ICT, display a positive attitude towards the technology of non-invasive brain-computer interfaces. They see some benefits, such as intuitive and inconspicuous operation and multiple areas of applications of such interfaces, most notably in medicine, e.g. for diagnostic purposes, emergency response, smart home technology and for active, ambient and assisted living. However, according to most of our participants, the benefits of using BCI are, for them, not significantly greater than issuing commands by voice. In particular, many of the participants claimed to prefer the concept of voice interaction to BCI in a smart home setting, which could be connected to their familiarity with the concept of voice interfaces. We also observed strong opposition to the design of a headband as a BCI smart home controller. It was deemed potentially uncomfortable and inferior to voice recognition which, in contrast, requires no wearable. Therefore, if used, we should put particular emphasis on making the BCI control device as discreet and non-invasive as possible. 

We are planning to continue workshop-based exploratory studies with larger groups of older adults, including those with different needs associated with underlying medical conditions. We would also like to evaluate attitudes towards different BCI control devices and the ethical factors associated with their use. There is also space for participatory design, especially of the feedback granted by the device while in training to use it or the preferred training simulation format, as well as the nature of the response to acknowledge the commands given.

\begin{acks}
We would like to thank older adults from our Living Lab, those affiliated with Kobo Association who participated in this study and all interdisciplinary experts involved with the HASE Lab group (Human Aspects in Software and System Engineering).
\end{acks}

\bibliographystyle{ACM-Reference-Format}
\bibliography{sample-base}

\appendix

\end{document}